\documentstyle[twoside,color,graphicx,fleqn,espcrc2]{article}

\def\abs#1{\left| #1\right|}
\def\mod#1{\abs{#1}}
\def\Im{\mathop{\mbox{Im} }}
\def\Re{\mathop{\mbox{Re} }}

\def\eprime{$\varepsilon'$}
\def\ratio{$\varepsilon'/\varepsilon$}

\definecolor{red}{rgb}{1,0,0}

\newcommand{\bea}{\begin{eqnarray}}
\newcommand{\be}{\begin{equation}}
\newcommand{\eea}{\end{eqnarray}}
\newcommand{\ee}{\end{equation}}
\newcommand{\nnu}{\nonumber}

\newcommand{\AmS}{{\protect\the\textfont2
  A\kern-.1667em\lower.5ex\hbox{M}\kern-.125emS}}

\title{Estimating \ratio\ . A user's manual}

\author{M.~Fabbrichesi\address{INFN, Sezione di Trieste and Scuola 
       Internazionale di Studi Superiori Avanzati,\\
       via Beirut 4, I-34014 Trieste, Italy.}%
       }%

\begin{document}

\pagestyle{empty}

\begin{abstract}
I review the current theoretical estimates of the CP-violating
parameter \ratio\ , compare
them to the experimental result and suggest a few guidelines in
using the theoretical results. 
\end{abstract}

\maketitle

\section{Notation}
\noindent
The parameter \eprime\ measures
direct CP violation and is defined by
the difference of the amplitude ratios
\bea
\varepsilon' & =  &
\frac{\varepsilon}{\sqrt{2}} \left\{
\frac{\langle ( \pi \pi )_{I=2} | {\cal L}_W | K_L \rangle}
{\langle ( \pi \pi )_{I=0} | {\cal L}_W | K_L \rangle} \right. \nnu \\
& & \quad \quad \quad \quad 
 - \left. \frac{\langle ( \pi \pi )_{I=2} | {\cal L}_W | K_S \rangle}
{\langle ( \pi \pi )_{I=0} | {\cal L}_W | K_S \rangle} \right\} 
\label{eps'} \, ,
\eea
where $K_{L,S}$ are the long- and short-lived neutral kaons, and
$\varepsilon$ measures indirect CP violation in the same system.

It is useful to recast eq.~(\ref{eps'}) in the form
\be
 \frac{\varepsilon '}{\varepsilon} = 
\frac{G_F \omega}{2\mod{\epsilon}\Re{A_0}} \:
\mbox{Im}\, \lambda_t  \:
 \left[ \Pi_0 - \frac{1}{\omega} \: \Pi_2 \right] \; ,
\label{eps'2}
 \ee
where, referring to the $\Delta S=1$ quark hamiltonian
 \be
 {\cal H}_{W}= \sum_i  \frac{G_F}{\sqrt{2}}  V_{ud}\,V^*_{us}
 \Bigl[z_i(\mu) + \tau\ y_i(\mu) \Bigr]  \; Q_i (\mu) \, ,
 \label{Lquark}
\ee
we have that
\bea
 \Pi_0 & = &  
\frac{1}{\cos\delta_0} \sum_i y_i \, \Re  \langle  Q_i  \rangle _0 
\ (1 - \Omega_{\eta +\eta'}) \; ,
\label{Pi_0}\\
 \Pi_2 & = & 
\frac{1}{\cos\delta_2} \sum_i y_i \, \Re \langle Q_i \rangle_2   
\label{Pi_2} \, ,
\eea
where the Wilson
coefficients $z_i$ and $y_i$ are known to the next-to-leading
order in $\alpha_s$ and $\alpha_e$~\cite{NLO}.
The four-quark operators $Q_{1 \cdots 10}$ are the standard set
 \be
\begin{array}{rcl}
Q_{1} & = & \left( \overline{s}_{\alpha} u_{\beta}  \right)_{\rm V-A}
            \left( \overline{u}_{\beta}  d_{\alpha} \right)_{\rm V-A}
\, , \\[1ex]
Q_{2} & = & \left( \overline{s} u \right)_{\rm V-A}
            \left( \overline{u} d \right)_{\rm V-A}
\, , \\[1ex]
Q_{3,5} & = & \left( \overline{s} d \right)_{\rm V-A}
   \sum_{q} \left( \overline{q} q \right)_{\rm V\mp A}
\, , \\[1ex]
Q_{4,6} & = & \left( \overline{s}_{\alpha} d_{\beta}  \right)_{\rm V-A}
   \sum_{q} ( \overline{q}_{\beta}  q_{\alpha} )_{\rm V\mp A}
\, , \\[1ex]
Q_{7,9} & = & \frac{3}{2} \left( \overline{s} d \right)_{\rm V-A}
         \sum_{q} \hat{e}_q \left( \overline{q} q \right)_{\rm V\pm A}
\, , \\[1ex]
Q_{8,10} & = & \frac{3}{2} \left( \overline{s}_{\alpha} 
                                                 d_{\beta} \right)_{\rm V-A}
     \sum_{q} \hat{e}_q ( \overline{q}_{\beta}  q_{\alpha})_{\rm V\pm A}
\, , 
\end{array}  
\label{Q1-10} 
\ee
and the hadronic matrix elements are
taken along the isospin direction $I=0$ and 2. Accordingly, $A_0$ is
the amplitude $A(K^0 \rightarrow \pi\pi, I=0)$ and $\omega$ is the
ratio $\Re A_2/\Re A_0$, the smallness of which goes under the name of
the $\Delta I=1/2$ rule; it plays an important role in the theoretical
prediction of \eprime\ . 
 The parameters $\tau \equiv - V_{td}
V_{ts}^{*}/V_{ud} V_{us}^{*}$ and  $\Im \lambda_t \equiv V_{td}\,V^*_{ts}$
are combinations of Cabibbo-Kobayashi-Maskawa coefficients.
See the review on \ratio\ in ref.~\cite{review} for the definition of the
isospin-breaking correction $\Omega_{\eta +\eta'}$ and the final-state
interaction phases $\delta_{0,2}$ as well as more details on the
definitions above. 

\section{Preliminary remarks}
\noindent
Let us go back to eq.~(\ref{eps'2}), where
\be
\frac{G_F \omega}{2 | \varepsilon | \,\mbox{Re}\, A_0} 
\simeq 10^{3} 
\: \mbox{GeV}^{-3} \, .
\ee
If  we were to take
\be
\Pi_{0,2}  
\simeq \frac{\alpha_s}{\pi} \, [m_K]^3 \simeq 10^{-2} 
\: \mbox{GeV}^{-3} \, ,
\ee
that is, estimating the hadronic matrix elements by simple dimensional
analysis ($\alpha_s/\pi$ takes into account the size of QCD
induced Wilson coefficients), and
\be
\mbox{Im}\, \lambda_t \simeq 10^{-4} \, ,
\ee
which is certainly reasonable, we would obtain
\be
\varepsilon'/\varepsilon \simeq 10^{-3} \, , \label{backenvelope}
\ee
a back-of-the-envelope estimate which is remarkably close to the
experimental result. What is then the problem? 
 
Consider  the contribution of the
various operators in the (very simple minded) vacuum saturation approximation
for the hadronic matrix elements.
Figure~1 visualizes this computation
in a pie chart that graphically shows how the contributions come with
different signs and that cancellations among them
can be sizable. 
 Dimensional analysis cannot be assumed to be reliable in the presence
of such large cancellations and therefore the result
(\ref{backenvelope}) cannot be trusted. 

\begin{figure}
\includegraphics[scale=0.4]{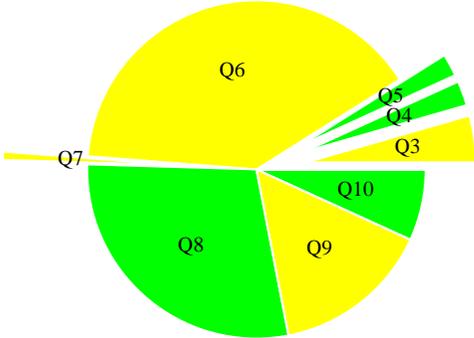}
\caption{The pie chart: In yellow/light-gray (green/dark-gray) 
positive (negative)
contributions to \ratio\ of the effective operators $Q_{1 \cdot
10}$. Hadronic matrix elements in the vacuum saturation approximation.}
\label{pie}
\end{figure}
 
The actual cancellation depends on the size of the hadronic matrix elements  
$\langle  Q_i \rangle_{0,2}$, the estimate of which
requires some control on the non-perturbative part of
QCD. This is by far the main source of uncertainty in any theoretical
estimate of \ratio\ . In addition, also the determination
of the overall factor $\Im \lambda_t$ depends on the non-perturbative 
amplitude for the transition of $K^0-\bar K^0$, thus making
the final uncertainty even larger.~\footnote{The presence of large
non-perturbative uncertainties is, in a nutshell, the reason
why \ratio\ is in general such a bad place where to look for new physics.}

It is on the basis of such a cancellation that,
in the early 90s, the idea that \ratio\ could be very small---of the
order of $O(10^{-4})$ if not
altogether vanishing (thus mimicking the super-weak scenario)---took
hold of the theoretical community.~\footnote{That idea was made stronger by the
ever-growing mass of the top quark that made such a cancellation 
between gluon and
electroweak penguin operators more and more effective.} At the same time, the
discrepancy between the two experimental results and in particular
the smallness of the FNAL result played a role in favor of
a small \ratio\ .

It is only this year (1999) that the (preliminary) results from the
new-generation
experiments have finally settled the question of the size of \ratio
and converged on the value
\be
\varepsilon'/\varepsilon = (2.1 \pm 0.46 ) \times 10^{-3} \label{exp}
\ee
which is obtained by averaging over the preliminary results
for the  1998-99 experiments
(KTeV~\cite{KTeV} and NA48~\cite{NA48}) as well as those in 
1992-93 (NA31~\cite{NA31} and E731~\cite{E731}). The result in 
eq.~({\ref{exp}) rules out the super-weak scenario and makes 
possible a detailed comparison with and among the theoretical analyses.

\section{Experiment vs. theoretical estimates}
\noindent
Given the fact that the gluon and electroweak penguin operator tend to
cancel each other contribution to \ratio\ , the question is whether
this cancellation is as effective as reducing by one order of
magnitude the
back-of-the-envelope estimate of the previous section or not.

Before the publication of this year experimental results, there were
three estimates of \ratio\ . Two of them (M\"unich and Rome) for
which the cancellation took place and one (Trieste) for which it did not.

The situation has not really changed this year, except for those new
estimates that have come out, partially confirming the Trieste prediction
of a large \ratio\ .

To compare different approaches, it is useful to introduce the
parameters
\be
B_i(\mu)^{(0,2)} = 
\frac{\langle Q_i \rangle_{0,2}}{\langle Q_i \rangle_{0,2}^{VSA}}
\ee
which give the correction in the approach with the respect to the result
in the vacuum saturation approximation (VSA). Let me stress that
there is nothing magical about the VSA: it is just a convenient (but arbitrary)
normalization point. Accordingly, there is no reason whatsoever to
prefer values of $B_i =1$ and most the cases in which the parameters
have been computed they have greatly deviated from 1---a case in point
is the parameter $B_1^{(0)}$,  which can be determined from the CP
conserving amplitude $A_0$ and is ten times bigger than its VSA
value because of the $\Delta I =1/2$ rule. 

{\scriptsize
\begin{table}
\begin{center}
\begin{tabular}{c|| c| c| c}
 & \textcolor{blue}{(ph)} &\textcolor{blue}{(lattice)} 
&\textcolor{blue}{($\chi$QM)} \\[1ex]
$B_i$ &      $\mu = 1.3$ GeV & $\mu = 2.0$ GeV &  $\mu = 0.8$ GeV\\[1ex]
\hline
& & &\\[1ex]
 $B_1^{(0)}$& 13~(\dag)  &  - &9.5  \\[1ex]
 $B_2^{(0)}$&$ 6.1\pm 1.0$~(\dag)  & - &2.9  \\[1ex]
 $B_1^{(2)} = B_2^{(2)}$& 0.48~(\dag)  &  - & 0.41 \\[1ex]
 $B_3$& 1~(*)  & 1~(*) & $-2.3$ \\[1ex]
 $B_4$&5.2~(*)  &$1 \div 6$~(*)  & 1.9 \\[1ex]
 \textcolor{red}{$B_5 \simeq B_6$}& \textcolor{red}{ $1.0\pm 0.3$~$(*)$} &
 \textcolor{red}{ $1.0\pm 0.2$~$(*)$} & \textcolor{red}{  $1.6\pm 0.3$} \\[1ex]
 $B_7^{(0)} \simeq B_8^{(0)}$&1~(*)   & 1~(*) & $2.5\pm 0.1$  \\[1ex]
 $B_9^{(0)}$& 7.0~(*\dag)  &1~(*)  & 3.6 \\[1ex]
 $B_{10}^{(0)}$& 7.5~(*\dag)  &1~(*)  & 4.4 \\[1ex]
 $B_7^{(2)}$&1~(*)  &$ 0.6\pm 0.1$  &  $0.92\pm 0.02$ \\[1ex]
 \textcolor{red}{$B_8^{(2)}$}& \textcolor{red}{ $0.8\pm 0.2$~$(*)$}  &
 \textcolor{red}{$ 0.8\pm 0.15$}  & \textcolor{red}{ $0.92\pm 0.02$} \\[1ex]
 $B_9^{(2)}$& 0.48   &$ 0.62\pm 0.10$  &0.41  \\[1ex]
 $B_{10}^{(2)}$&0.48   & 1~(*) & 0.41 \\[1ex]
 \textcolor{red}{$B_K$}& \textcolor{red}{$ 0.80\pm 0.15$}   &
 \textcolor{red}{$ 0.75\pm 0.15$}  & \textcolor{red}{$ 1.1\pm 0.2$}  \\[1ex]
& & &\\[1ex]
\hline
\end{tabular}
\caption{The $B_i$ factors in three approaches. ($\dag$) stands
for an input value and (*) for an ``educated guess''.}
\end{center}
\end{table}
}
 
Table 1 collects the $B_i$ parameter for three approaches. 
Notice that larger values of $B_K$ give smaller values for
$\Im \lambda_t$ and accordingly for \ratio\ . In discussing the
various approaches, it is important to
bear in mind that a $B_i$ parameter, being normalized on the VSA,
could depend on a quantity, like $m_s$, even when the
estimate itself does not.

Let consider
the two most relevant operators $B_6$ and $B_8$. There is overall
agreement
among the various approaches on $B_8$. On the other hand, for the
crucial parameter $B_6$ the M\"unich and Roma group must relay on an
``educated guess'' and only the Trieste group provides a computed
value. For this reason, I think that is fair to say
that both the M\"unich and Roma~\footnote{See~\cite{latticeb6} for comments
about the unreliability of the previous lattice estimate of $Q_6$.} 
estimate suffer of a systematic
bias in so far as the crucial parameter $B_6$ is not estimated but
simply varied around the large $1/N_c$ (vacuum saturation) result. 
In a computation
that essentially consists in the difference between two contributions,
the fact that one of the two is simply assumed to vary around a
completely arbitrary central value casts some doubts about any
statement about unlikely corners of parameter space for which the
current experimental result  can be reproduced by the
theory.

\section{Extended caption to Fig.~\ref{expvsth}}
\noindent
The simplest way of summarizing the present status of theoretical estimates
of \ratio\ consists in explaining Fig.~\ref{expvsth}. Let us group the
various estimated according on whether they were published before or
after the last run of experiments (early 1999), in other words, between
those published when the value of \ratio\ was still uncertain and
those after it has been determined to be $\approx 2 \times
10^{-3}$. The
various approaches substantially agree on the short-distance analysis and
inputs and therefore I will only discuss here the long-distance
part.~\footnote{Most current estimates, in trying to reduce the final
error,  treat the uncertainties of the
experimental inputs via a Gaussian distribution as opposed to a flat
scanning. }  
\clearpage
\begin{itemize}
\item Pre-dictions:
\begin{itemize}
\item \fbox{Roma 1996~\cite{roma96}} It is based on the lattice 
simulation of non-perturbative QCD. The hadronic matrix elements
are included using the lattice simulation for those known and
``educated guesses'' for those which are not known. 
Only the Gaussian treatment of the uncertainties (red/dark-gray bar) is given.

\item \fbox{M\"unich 1996~\cite{munich96}} It is based on a mixture of
phenomenological and $1/N_c$ approach in which as many as possible of
the matrix elements are determined by means of known CP conserving
amplitudes and those remaining by leading $1/N_c$ estimates. 
Both the flat scanning (light-blue/light-gray) and the
Gaussian treatment (red/dark-gray) 
of the uncertainties is given. The two values correspond to
two different choices for the strange quark mass.

\item \fbox{Trieste 1997~\cite{trieste97}} It is based on the chiral quark
model. All matrix elements are parameterized in terms of three
parameters: the quark and gluon condensates and the constituent quark
mass. The values of these parameters
are determined by fitting the $\Delta I =1/2$ rule.
Chiral perturbation corrections are included to the complete $O(p^4)$.
Both the flat scanning (light-blue/light-gray) and the
Gaussian
treatment (red/dark-gray) of 
the uncertainties is given.~\footnote{I thank F.~Parodi for the
Gaussian estimate of the error in the chiral-quark model result.}
\end{itemize}

\item Post-dictions

\begin{itemize}
\item \fbox{M\"unich 1999~\cite{munich99}} It is an updated analysis similar to that
of 1996. The two ranges are now those obtained by using the Wilson
coefficients in the HV and NDR regularization prescription. Again,
both the flat scanning (light-blue/light-gray) and the
Gaussian
treatment (red/dark-gray) of the uncertainties is given.

\item \fbox{Dortmund 1999~\cite{dortmund}} It is based on the 
$1/N_c$ estimate of the
hadronic matrix elements, regularized by means of an explicit cutoff.
Chiral perturbation corrections are included (partially) up to
$O(p^4)$. Two estimates are given according to whether the input
parameters are kept fixed (red/dark-gray) or varied
(light-blue/light-gray). The second range given corresponds to the
inclusion of important $O(p^4)$ corrections.
No central values are given.

\item \fbox{Dubna 1999~\cite{dubna}} It is based on chiral 
perturbation theory up
to  $O(p^6)$. I cannot say much about it because
it came out just at the time of this conference. I have
included their full range according to the tables
reported in the reference above (flat scanning in
light-blue/light-gray, the Gaussian treatment in red/dark-gray).
\end{itemize}
\end{itemize}
\begin{figure}
\includegraphics[scale=0.55]{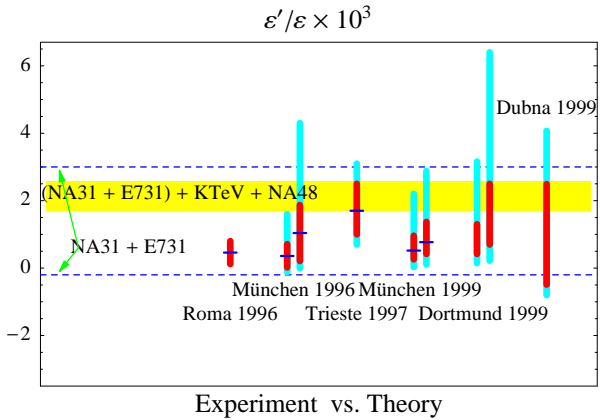}
\caption{Experiment vs. theoretical estimates. See text for full caption.}
\label{expvsth}
\end{figure}
\section{Strengths and weaknesses of the various approaches}
\noindent
Since there is no estimate which is safe from criticism, 
I would like to try to summarize the strengths and weaknesses of the
various approaches and leave it to the reader to decide by himself.

\begin{itemize}
\item \fbox{Roma}\begin{itemize}
\item Good: The lattice approach is well-grounded in first-principles.
\item Bad: Half of the computation is missing: there is no
determination of $B_6$, the value of which must be guessed. 
\end{itemize}

\item \fbox{M\"unich}\begin{itemize}
\item Good: Clever use of CP conserving amplitudes. Determination of
many $B_i$ in a model-independent manner.
\item Bad: The important parameters 
$B_8$ and $B_6$ cannot be determined and must be varied around their
leading $1/N_c$ values.
\end{itemize}

\item \fbox{Trieste}\begin{itemize}
\item Good: All operators ($Q_6$ included) are
determined in a consistent manner;
the full $O(p^4)$ chiral
perturbation is included. 
\item Bad: Phenomenological model which is not derivable
from first principle. There is a 
uncertainty in the matching procedure which is  difficult to
estimate.
\end{itemize}

\item \fbox{Dortmund}\begin{itemize}
\item Good:  State-of-the-art $1/N_c$ estimate of all matrix elements.
\item Bad: Potentially important $O(p^4)$ 
not included yet in the analysis. Unstable matching and
therefore very large uncertainties.
\end{itemize}
\end{itemize}

\section{The lesson of the chiral quark model result}
\noindent
The crucial enhancement of the parameter $B_6$ in the chiral quark
model originates in the fit of the $\Delta I =1/2$ rule that is at the
basis of this model. We could say that it is a revival of the old 
idea~\cite{VG}
of having the same gluon penguin
operator explaining the  $\Delta I =1/2$ also give
a large \ratio\ (see Fig.~4). 
This mechanism works only at a scale around 1 GeV and is
not as complete as in the original idea (in the chiral quark model the
penguin contribution to the $A_0$ amplitude turns out to be about
20\%, see Fig.~3). Clearly for approaches based on scales higher than $m_c$
 a different mechanism must be at work to mimic the same effect 
(given the scale
invariance of the physical amplitudes).
\begin{figure}
\includegraphics[scale=0.4]{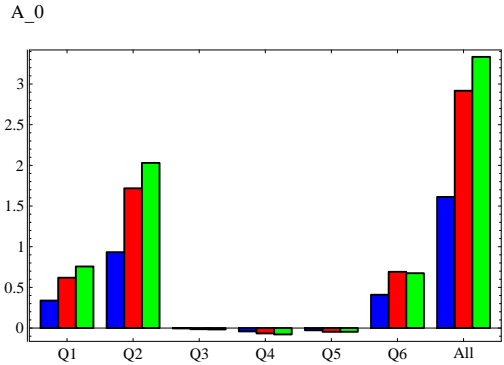}
\caption{The amplitude $A_0$ in the chiral quark model. The
contribution of the gluon penguin operator to order $O(p^4)$
(green/light-gray histogram) is about 20\% of the total.}
\label{chart0}
\end{figure}
\begin{figure}
\includegraphics[scale=0.4]{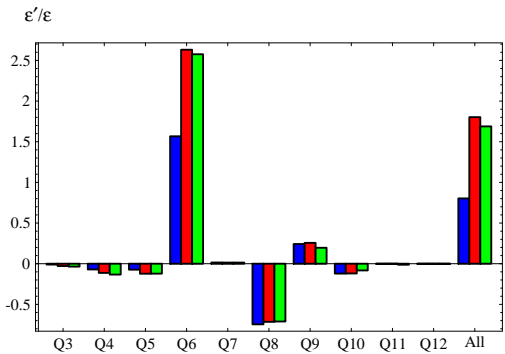}
\caption{Anatomy of \ratio\ in the chiral quark model. The ratio is
dominated by the gluon penguin operator $Q_6$ but only after all
correction order  $O(p^4)$ are included (the green/light-gray histogram).}
\label{isto_qm}
\end{figure}

\section{Conclusion}
\noindent
As Fig.~2 makes it clear, there is no disagreement between the
experimental result and the prediction of the standard model
once all uncertainties are properly taken into account. Of the
five available estimates today (August 1999), 
three~\footnote{Even though it is true that two of them (Dortmund and
Dubna) suffer of very large errors and can only be used as indications
rather than real estimates.} overlap with the experimental
range and one of them (Trieste) even predicted it two years in advance
of the experiments.~\footnote{It is particularly remarkable that the only
prediction that eventually agreed with the experiment turned out to be
also the only one that estimated (albeit within a phenomenological model) all
hadronic matrix elements and satisfied the $\Delta I=1/2$ rule.} 
Only the Rome and M\"unich estimates are somewhat
below the current experimental range but they suffer of a systematic
uncertainty, as discussed in the previous section.

If we abstract from the details and the
central values of the various estimates, it is comforting that in such
a complicated computation, different approaches give
results that are rather consistent  among themselves (namely, values of
\ratio\ positive and of the order of $O(10^{-3})$) and
in overall agreement with the experiment. This, I think,
is the most important message.  

A final word on possible future improvements. 

The place where to look for a reduction of the
present theoretical uncertainties is $\Im \lambda_t$. Ideally, this
coefficient could be determined in a manner that is free of
non-perturbative uncertainties in the process $K_L\to\pi^0 \bar \nu
\nu$. Such a determination could easily reduce the uncertainty in \ratio\
by 20-30\%.

On the front of hadronic matrix elements, work is in progress
on various phenomenological approaches as well as on lattice simulations.

\bigskip\bigskip
{\it
{\sc Question ({\bf M. Neubert}, SLAC):} 
What is the basis for the ``educated
guesses''
leading to values of $B_{6,8}$ close to one, given that all the other
B-parameters known from data show very large deviations form the
vacuum saturation approximation?

{\sc Answer:} My very same objection.
Anyway, some come from leading $1/N_c$ estimates, some are just
guesses. 

{\sc Question ({\bf L. Giusti}, Boston Univ.):} 
Which $B_K$ do you use to
extract $\Im \lambda_t$ in the chiral quark model?

{\sc Answer:} That determined in the chiral quark model, see
Table~1. As a matter of fact, it is because in this model $B_K$ comes
out larger than in other estimates that \ratio\ is not even larger.
}


\bigskip\medskip

\begin{thebibliography}{99}

\bibitem{review} {\sc S. Bertolini, J. Eeg and M. Fabbrichesi},
{\tt hep-ph/9802405}, to appear in
Reviews of Modern Physics, January 2000.

\bibitem{NLO} {\sc A.~J.~Buras et al}, Nucl. Phys. B370 (1992) 69 and 
 Nucl. Phys. B400 (1993) 37;\\
{\sc A.~J.~Buras, M. Jamin and M.E. Lautenbacher},  Nucl. Phys. B400
(1993) 75 and  Nucl. Phys. B408 (1993) 209;\\
{\sc M. Ciuchini, E. Franco, G. Martinelli and L. Reina}, 
 Nucl. Phys. B415 (1994) 403.

\bibitem{KTeV} {\sc A. Alavi-Harati et~al.}, Phys. Rev. Lett. 83
(1999) 22.

\bibitem{NA48} {\sc P.~Debu}, seminar at CERN, June 18, 1999 ({\tt
http://www.cern.ch/NA48}).

\bibitem{NA31} {\sc G.~D.~Barr et~al.}, Phys. Lett. B317 (1993) 233.

\bibitem{E731} {\sc L.~K. Gibbons et~al.}, Phys. Rev. D55 (1997) 6625.

\bibitem{latticeb6} R. Gupta, {\tt hep-ph/9801412};\\ 
{G. Martinelli,\tt hep-lat/9810013};\\
{\sc D. Pekurovsky and G. Kilcup}, {\tt hep-lat/9812019}.
 
\bibitem{roma96} {\sc M. Ciuchini}, Nucl. Phys. B Proc. Suppl. 59
(1997) 149.

\bibitem{munich96} {\sc A.~J. Buras, M. Jamin and M.~E. Leutenbacher},
Phys. Lett. B389 (1996) 749. 

\bibitem{trieste97} {\sc S. Bertolini et al.}, Nucl. Phys. B514 (1998) 93. 

\bibitem{munich99} {\sc S. Bosch et al.}, {\tt hep-ph/9904408}

\bibitem{dortmund} {\sc T. Hambye et al.}, {\tt hep-ph/9906434}.

\bibitem{dubna} {\sc A.~A. Bel'kov et al. },  {\tt hep-ph/9907335}.

\bibitem{VG} {\sc A. I. Vainshtein et al.}, JEPT 45 (1977) 670;\\ 
{\sc F. J. Gilman and M. B. Wise}, Phys. Lett. B83 (1979)83;\\
{\sc B. Guberina and R. D. Peccei}, Nucl. Phys. B163 (1980) 289.


\end{thebibliography}
\end{document}